# Microsimulation Estimates of Decision Uncertainty and Value of Information Are Biased but Consistent

Jeremy D. Goldhaber-Fiebert, PhD (1,2), Hawre Jalal, MD, PhD (3), Fernando Alarid Escudero, PhD (1,2)

(1) Department of Health Policy, Stanford School of Medicine, Stanford, CA, USA
(2) Center for Health Policy, Freeman Spogli Institute, Stanford University, Stanford, CA, USA
(3) School of Epidemiology and Public Health, University of Ottawa, Ottawa, Ontario, Canada

**Corresponding Author:**
Jeremy D. Goldhaber-Fiebert
Department of Health Policy
Stanford University
615 Crothers Way
Stanford, CA, 94305
jeremygf@stanford.edu
650-721-2486

**Word Count:** Currently about 4,530

**Abstract (275 words)**

**Purpose:** Individual-level state-transition microsimulations (iSTMs) have proliferated for economic evaluations in place of cohort state transition models (cSTMs). Probabilistic economic evaluations quantify decision uncertainty and value of information (VOI). Previous studies show that iSTMs provide unbiased estimates of expected incremental net monetary benefits (EINMB), but statistical properties of iSTM-produced estimates of decision uncertainty and VOI remain uncharacterized.

**Methods:** We compare iSTMs-produced estimates of decision uncertainty and VOI to corresponding cSTMs. For a 2-alternative decision and normally distributed incremental costs and benefits, we derive analytical expressions for the probability of being cost-effective and the expected value of perfect information (EVPI) for cSTMs and iSTMs, accounting for correlations in incremental outcomes at the population and individual levels. We use numerical simulations to illustrate our findings and explore the impact of relaxing normality assumptions or having >2 decision alternatives.

**Results:** iSTM estimates of decision uncertainty and VOI are biased but asymptotically consistent (i.e., bias approaches 0 as number of microsimulated individuals approaches infinity). Decision uncertainty depends on one tail of the INMB distribution (e.g., P(INMB <0)) which depends on estimated variance (larger with iSTMs given first-order noise). While iSTMs overestimate EVPI, their direction of bias for the probability of being cost-effective is ambiguous. Bias is larger when uncertainties in incremental costs and effects are negatively correlated since this increases INMB variance.

**Conclusions:** iSTMs are useful for probabilistic economic evaluations. While more samples at the population uncertainty level are interchangeable with more microsimulations for estimating EINMB, minimizing iSTM bias in estimating decision uncertainty and VOI depends on sufficient microsimulations. Analysts should account for this when allocating their computational budgets and, at minimum, characterize such bias in their reported results.

**Highlights**

Individual-level state-transition microsimulation models (iSTMs) produce biased but consistent estimates of the probability that interventions are cost-effective.

iSTMs also produce biased but consistent estimates of the expected value of perfect information.

The biases in these decision uncertainty and value of information measures are not reduced by more parameter sets being sampled their population-level uncertainty distribution but rather by more individuals being microsimulated for each parameter set sampled.

Analysts using iSTMs to quantify decision uncertainty and value of information should account for these biases when allocating their computational budgets and, at minimum, characterize such bias in their reported results.

**Introduction**

Model-based economic evaluations are increasingly used to inform decisions about which interventions to implement and whether there is value in conducting potential future research. The use of probabilistic analyses within economic evaluations has gained increasing traction and is now the recommended approach for base-case analyses.[1-3] Probabilistic analyses permit unbiased estimates of expected outcomes for non-linear models and quantification of decision uncertainty stemming from uncertainty in population-level model inputs (i.e., second-order parameter uncertainty). Because uncertainty in model parameters can be correlated, prior work has developed approaches for estimating, sampling from, and propagating the joint uncertainty distribution of model parameters.[4-11]

While cohort state-transition models (cSTMs) (i.e., mean-process models) were the standard modeling technique in the past, individual-level state-transition microsimulations (iSTMs) are increasingly used for economic evaluations.[12,13] For any cSTM, a corresponding and equivalent iSTM that models the same underlying processes can be formulated.[12] The expected outcomes produced by simulations using an iSTM are estimates of the corresponding mean-process outcomes obtained from simulations with the equivalent cSTM.[14]

Several trends in medicine and public health have driven the increased use of iSTMs. These include: 1) the proliferation of high-performance computing[10,12]; 2) the focus on sophisticated and personalized interventions coupled with the rise of sophisticated risk prediction models that depend on a complex set of dynamic individual characteristics[15,16]; and 3) the move to proactively address health disparities, with health equity defined across multiple dimensions (e.g., race/ethnicity, socioeconomic status, and social determinants of health).[17-19]

When a sufficiently large number of individuals are simulated using an iSTM, the outcomes produced converge to those that would have been produced by a corresponding cSTM.[12,20-23] However, when smaller numbers of individuals are simulated, modeled outcomes produced by the iSTM can differ appreciably from the cSTM's outcomes due to Monte Carlo noise (i.e., first-order uncertainty). Prior work conceptualizes the rate of convergence and, hence, the necessary number of microsimulated individuals needed to estimate expected outcomes and expected incremental outcomes within a given error bound.[20-23]

Prior work also considers probabilistic analyses using iSTMs,[21-23] which involves both first-order and second-order uncertainty. Specifically, this prior work developed methods to produce estimates with as few samples as possible from the joint model parameter uncertainty distribution because each sample then involves computationally costly microsimulations of many individuals. A key finding is that for a given number of samples from the joint model parameter uncertainty distribution, estimates of outcomes made with an iSTM will be unbiased but that their uncertainty will be systematically overestimated because they also contain Monte Carlo noise.

While the prior work cited here has focused on making estimates of expected outcomes and expected incremental outcomes providing guidance on the number of samples to take from the joint uncertainty distribution and number of individuals to microsimulate, it does not provide guidance for other quantities of interest related to decision uncertainty and the value of information. The present study provides analytic results and numerical simulations to characterize the precision and bias of estimating such quantities using microsimulation models.

**Theoretical Model**

We explore the precision and bias of estimates of decision uncertainty and the value of information using a simple decision model with two alternatives (treatment vs. no treatment). We make these estimates using an iSTM with various numbers of sampled individuals ($n_{micro}$) and compare them to estimates produced with a cSTM (i.e., an iSTM where $n_{micro} = \infty$).

*Cohort State-Transition Markov Model*

Let $\Theta = \{\theta_{\Delta Q}, \theta_{\Delta C}\}$ be the parameters for the incremental QALYs and Costs of the decision model with $\{\sigma_{\Delta Q}, \sigma_{\Delta C}\}$ reflecting the current level of second-order uncertainty regarding these incremental values.

We first specify the cSTM. At the population level, we assume that the uncertainty in the incremental discounted lifetime QALYs with treatment compared to with no treatment is characterized by:

$$\Delta Q_{psa} \sim N(\theta_{\Delta Q}, \sigma_{\Delta Q}) \quad (1)$$

While the random variables like $\Delta Q_{psa}$ that we will consider below need not be normally distributed, we make this assumption because doing so will permit the analytic expressions we derive later to be more concrete. Just as with the incremental QALYs, at the population level, we assume that the uncertainty in the incremental discounted lifetime Costs with treatment compared to with no treatment is characterized by:

$$\Delta C_{psa} \sim N(\theta_{\Delta C}, \sigma_{\Delta C}) \quad (2)$$

More generally, as the uncertainty of incremental QALYs and incremental Costs may be correlated, we can rewrite their joint uncertainty as:

$$\langle \Delta Q_{psa}, \Delta C_{psa} \rangle \sim MVN(\langle \theta_{\Delta Q}, \theta_{\Delta C} \rangle, \Sigma_{psa}), \quad (3)$$

where $\Sigma_{psa}$ is the variance-covariance matrix, where the diagonal consists of $\sigma^2_{\Delta Q}$ and $\sigma^2_{\Delta C}$, and the off-diagonal elements are defined as $\rho_{psa}\sigma_{\Delta Q}\sigma_{\Delta C}$, with correlation in the population uncertainty of incremental QALYs and incremental Costs equal to $\rho_{psa}$.

For the cSTM, a given draw from the population distribution of incremental QALYs and incremental Costs (i.e., PSA sample) can be expressed in terms of incremental net monetary benefit:

$$INMB_{psa} = \lambda \Delta Q_{psa} - \Delta C_{psa} \quad (4)$$

where $\lambda$ is the willingness to pay for QALYs gained. The variance of the incremental net monetary benefit across PSA samples is then:

$$\begin{aligned} Var(INMB_{psa}) &= Var(\lambda \Delta Q_{psa} - \Delta C_{psa}) \\ &= \lambda^2 Var(\Delta Q_{psa}) + Var(\Delta C_{psa}) - 2\lambda Cov(\Delta Q_{psa}, \Delta C_{psa}) \\ &= (\lambda \sigma_{\Delta Q})^2 + (\sigma_{\Delta C})^2 - 2\lambda \rho_{psa} \sigma_{\Delta Q} \sigma_{\Delta C} \end{aligned} \quad (5)$$

Note that Equation (5), the variance of $INMB_{psa}$, does not rely on the normality assumptions above.
For the cSTM, the expected incremental net monetary benefit of treatment is:

$$EINMB_{psa} = E_\Theta(INMB_{psa}) \tag{6}$$

As the number of PSA samples is finite ($n_{psa}$), there is uncertainty in this unbiased estimate of $EINMB_{psa}$, given by

$$\begin{aligned} Var(EINMB_{psa}) &= \frac{Var(INMB_{psa})}{n_{psa}-1} \\ &= \frac{(\lambda\sigma_{\Delta Q})^2 + (\sigma_{\Delta C})^2 - 2\lambda\rho_{psa}\sigma_{\Delta Q}\sigma_{\Delta C}}{n_{psa}-1} \\ &= \lambda^2(se_{\Delta Q})^2 + (se_{\Delta C})^2 - 2\lambda\rho_{psa}se_{\Delta Q}se_{\Delta C}, \end{aligned} \tag{7}$$

where $se_{\Delta Q}$ and $se_{\Delta C}$ are standard errors for the estimated incremental QALYs and Costs, respectively.

The variance of $EINMB_{psa}$ in Equation (7) does not rely on the normality assumptions above.

In terms of decision uncertainty, i.e., the probability that a given strategy is the most cost-effective (often represented as a cost-effectiveness acceptability curve, or CEAC), the probability that treatment is the optimal strategy for the cSTM is:

$$\begin{aligned} CEAC_{psa} &= P(INMB_{psa} > 0) \\ &= 1 - \Phi\left(\frac{0 - EINMB_{psa}}{\sqrt{Var(INMB_{psa})}}\right) \\ &= 1 - \Phi\left(\frac{0 - EINMB_{psa}}{\sqrt{(\lambda\sigma_{\Delta Q})^2 + (\sigma_{\Delta C})^2 - 2\lambda\rho_{psa}\sigma_{\Delta Q}\sigma_{\Delta C}}}\right) \end{aligned} \tag{8}$$

where $\Phi(.)$ is the cumulative density function of the standard normal distribution.

The expected value of perfect information (EVPI) for the cSTM is defined as:

$$EVPI_{psa} = \begin{cases} E\left(max(0, -INMB_{psa})\right) & \text{if } EINMB_{psa} > 0 \\ E\left(max(0,\ INMB_{psa})\right) & \text{if } EINMB_{psa} < 0 \end{cases} \tag{9}$$

We require both cases because, at some values of $\lambda$, the expected incremental net monetary benefit of treatment ($EINMB_{psa}$) could be negative, implying that the preferred strategy on expectation is no treatment. This, in turn, implies that information will be valuable for PSA samples where the incremental benefit of treatment ($INMB_{psa}$) is positive.

In the case of $EINMB_{psa} > 0$, we can express the value of information as:

$$EVPI_{psa} = \int_{-\infty}^{0} \left(-INMB_{psa} \frac{1}{\sqrt{Var(INMB_{psa})}} \varphi\left(\frac{INMB_{psa} - EINMB_{psa}}{\sqrt{Var(INMB_{psa})}}\right)\right) dINMB_{psa} \tag{10}$$

where $\varphi(.)$ is the probability density function of the standard normal distribution. This expression is a form of the Unit Normal Loss Integral function which was first discussed by Raiffa and Schlaifer in the 1960s.[24-26] Likewise, in the case of $EINMB_{psa} < 0$, this expression is:

$$EVPI_{psa} = \int_0^\infty \left( INMB_{psa} \frac{1}{\sqrt{Var(INMB_{psa})}} \varphi \left( \frac{INMB_{psa} - EINMB_{psa}}{\sqrt{Var(INMB_{psa})}} \right) \right) dINMB_{psa} \quad (11)$$

*Individual-level Microsimulation Model*

If an iSTM equivalent to the cSTM is used to estimate the quantities described for the cSTM above using a given number of microsimulated individuals ($n_{micro}$), we arrive at the corresponding set of equations.

For any PSA sample from the population distribution, the distribution of incremental QALYs with treatment compared to with no treatment that is estimated from an iSTM is:

$$\Delta Q_{micro} \sim N(\Delta Q_{psa}, se_{\Delta Q, micro}) \text{ where } se_{\Delta Q, micro} = \frac{\sigma_{\Delta Q, micro}}{\sqrt{(n_{micro} - 1)}} \quad (12)$$

where $\Delta Q_{psa}$ is the population incremental QALYs with treatment sampled from the second-order uncertainty distribution and $\sigma_{\Delta Q, micro}$ is the square root of the variance of incremental QALYs with treatment across individuals in the microsimulation. For our simple example, we assume that the first-order noise from the microsimulations is homoscedastic with respect to the level of incremental QALYs with treatment, i.e., that the variance across microsimulated individuals in terms of incremental QALYs from treatment ($\sigma_{\Delta Q, micro}$) is independent of $\Delta Q_{psa}$. Hence, Equation (12) can then be expressed equivalently as:

$$\Delta Q_{micro} = \Delta Q_{psa} + \varepsilon_{\Delta Q, micro} \text{ where } \varepsilon_{\Delta Q, micro} \sim N(0, se_{\Delta Q, micro}) \quad (13)$$

In other words, an estimate of the population incremental QALYs generated from an iSTM will have the same expectation as the value produced by the cSTM with the addition of zero-centered noise due to the finite number of simulated individuals in the iSTM.

Likewise, for any PSA sample from the population distribution, the distribution of incremental Costs with treatment compared to with no treatment estimated from an iSTM is:

$$\Delta C_{micro} \sim N(\Delta C_{psa}, se_{\Delta C, micro}) \text{ where } se_{\Delta C, micro} = \frac{\sigma_{\Delta C, micro}}{\sqrt{(n_{micro} - 1)}} \quad (14)$$

For Equation (14) the assumptions are parallel to those for Equation (12) regarding incremental QALYs, which we then express as:

$$\Delta C_{micro} = \Delta C_{psa} + \varepsilon_{\Delta C, micro} \text{ where } \varepsilon_{\Delta C, micro} \sim N(0, se_{\Delta C, micro}) \quad (15)$$

Since the uncertainty of incremental QALYs and incremental Costs may be correlated within an individual in the microsimulation (i.e., some individuals could gain more than average number of QALYs with treatment and also have higher than average incremental costs with treatment in the case of positive within-individual correlation), we can rewrite the joint distribution as:

$$\langle \Delta Q_{micro}, \Delta C_{micro} \rangle \sim MVN\left(\langle \Delta Q_{psa}, \Delta C_{psa} \rangle, \Sigma_{micro}\right) \qquad (16)$$

where $\Sigma_{micro}$ is the variance-covariance matrix with diagonal elements defined by $(se_{\Delta Q,micro})^2$ and $(se_{\Delta C,micro})^2$ and off-diagonal elements defined as $\rho_{micro} se_{\Delta Q,micro} se_{\Delta C,micro}$, with correlation in the incremental QALYs and incremental costs experienced by individuals equal to $\rho_{micro}$.

For the iSTM, given a draw from the population distribution of incremental QALYs and incremental Costs (i.e., PSA sample), the miscrosimulated estimate of the incremental net monetary benefit can be expressed as:

$$INMB_{micro} = \lambda \Delta Q_{micro} - \Delta C_{micro}$$
$$INMB_{micro} = \lambda(\Delta Q_{psa} + \varepsilon_{\Delta Q,micro}) - (\Delta C_{psa} + \varepsilon_{\Delta C,micro}) \qquad (17)$$

where $\lambda$ is the willingness to pay for QALYs gained. The variance of the miscrosimulated incremental net monetary benefit across PSA samples is then:

$$
\begin{aligned}
Var(INMB_{micro}) &= Var(\lambda \Delta Q_{micro} - \Delta C_{micro}) \\
&= Var\left(\lambda(\Delta Q_{psa} + \varepsilon_{\Delta Q,micro}) - (\Delta C_{psa} + \varepsilon_{\Delta C,micro})\right) \\
&= \lambda^2 Var(\Delta Q_{psa}) + \lambda^2 Var(\varepsilon_{\Delta Q,micro}) + Var(\Delta C_{psa}) \\
&\quad + Var(\varepsilon_{\Delta C,micro}) - 2\lambda Cov(\Delta Q_{psa}, \Delta C_{psa}) \\
&\quad - 2\lambda Cov(\varepsilon_{\Delta Q,micro}, \varepsilon_{\Delta C,micro}) \\
&= (\lambda \sigma_{\Delta Q})^2 + (\sigma_{\Delta C})^2 + (\lambda se_{\Delta Q,micro})^2 + (se_{\Delta C,micro})^2 \\
&\quad - 2\lambda \rho_{psa} \sigma_{\Delta Q} \sigma_{\Delta C} - 2\lambda \rho_{micro} se_{\Delta Q,micro} se_{\Delta C,micro} \\
&= Var(INMB_{psa}) \\
&\quad + \left((\lambda se_{\Delta Q,micro})^2 + (se_{\Delta C,micro})^2 - 2\lambda \rho_{micro} se_{\Delta Q,micro} se_{\Delta C,micro}\right)
\end{aligned} \qquad (18)
$$

This expression for the variance of $INMB_{micro}$ does not rely on the normality assumptions above. Note that for simplicity, in the example we are considering here, we assume that the covariances between the microsimulation noise terms ($\varepsilon_{\Delta Q,micro}$ and $\varepsilon_{\Delta C,micro}$) and the PSA sampled quantities ($\Delta Q_{psa}$ and $\Delta C_{psa}$) are 0. In principle, if this were not the case, then we would have additional terms in the variance equation of the form $-2\rho_{X,Y} XY$ which when the correlation was less than 0 would lead to even larger variances. Note from Equation (18) that as the number of microsimulated individuals increase, i.e., $n_{micro} \to \infty$, $Var(INMB_{micro}) \to Var(INMB_{psa})$.

The expected INMB estimated with the iSTM, where expectations are taken over the parameter uncertainty ($\Theta$), is then:

$$EINMB_{micro} = E_\Theta(INMB_{micro}) \qquad (19)$$

$$\begin{aligned}
&= E\left(\lambda(\Delta Q_{psa} + \varepsilon_{\Delta Q,micro}) - (\Delta C_{psa} + \varepsilon_{\Delta C,micro})\right) \\
&= \lambda E(\Delta Q_{psa} + \varepsilon_{\Delta Q,micro}) - E(\Delta C_{psa} + \varepsilon_{\Delta C,micro}) \\
&= \lambda\left(E(\Delta Q_{psa}) + E(0)\right) - E(\Delta C_{psa}) - E(0) \\
&= E(\lambda \Delta Q_{psa} - \Delta C_{psa}) \\
&= E(INMB_{psa}) \\
&= EINMB_{psa}
\end{aligned}$$

Note that the final equation above implies that expected incremental net monetary benefit estimated with the iSTM is equal to the expected incremental net monetary benefit estimated with the cSTM.

The uncertainty of the expected incremental net monetary benefit estimated with the microsimulation over the PSA samples is:

$$\begin{aligned}
Var(EINMB_{micro}) &= \frac{Var(INMB_{micro})}{n_{psa} - 1} \\
&= \frac{(\lambda\sigma_{\Delta Q})^2 + (\sigma_{\Delta C})^2 + (\lambda se_{\Delta Q,micro})^2 + (se_{\Delta C,micro})^2 - 2\lambda\rho_{psa}\sigma_{\Delta Q}\sigma_{\Delta C} - 2\lambda\rho_{micro}se_{\Delta Q,micro}se_{\Delta C,micro}}{n_{psa} - 1} \\
&= \lambda^2(se_{\Delta Q})^2 + (se_{\Delta C})^2 - 2\lambda\rho_{psa}se_{\Delta Q}se_{\Delta C} \\
&\quad + \left(\frac{\lambda^2(se_{\Delta Q,micro})^2 + (se_{\Delta C,micro})^2 - 2\lambda\rho_{micro}se_{\Delta Q,micro}se_{\Delta C,micro}}{n_{psa} - 1}\right) \\
&= Var(EINMB_{psa}) + \left(\frac{\lambda^2(se_{\Delta Q,micro})^2 + (se_{\Delta C,micro})^2 - 2\lambda\rho_{micro}se_{\Delta Q,micro}se_{\Delta C,micro}}{n_{psa} - 1}\right)
\end{aligned} \quad (20)$$

This expression for the variance of $EINMB_{micro}$ in Equation (20) does not rely on the normality assumptions above.

In terms of decision uncertainty and the value of information, the probability that treatment is the optimal strategy for the iSTM is:

$$\begin{aligned}
CEAC_{micro} &= P(\lambda(\Delta Q_{psa} + \varepsilon_{\Delta Q,micro}) - E(C_{psa} + \varepsilon_{\Delta C,micro}) > 0) \\
&= 1 - \Phi\left(\frac{0 - EINMB_{micro}}{\sqrt{Var(INMB_{micro})}}\right) \\
&= 1 - \Phi\left(\frac{0 - EINMB_{micro}}{\sqrt{(\lambda\sigma_{\Delta Q})^2 + (\sigma_{\Delta C})^2 + (\lambda se_{\Delta Q,micro})^2 + (se_{\Delta C,micro})^2 - 2\lambda\rho_{psa}\sigma_{\Delta Q}\sigma_{\Delta C} - 2\lambda\rho_{micro}se_{\Delta Q,micro}se_{\Delta C,micro}}}\right)
\end{aligned} \quad (21)$$

The expected value of perfect information for the iSTM is defined as:

$$EVPI_{micro} = \begin{cases} E(max(0, -INMB_{micro})) & \text{if } EINMB_{micro} > 0 \\ E(max(0, \ INMB_{micro})) & \text{if } EINMB_{micro} < 0 \end{cases} \quad (22)$$

In the case of $EINMB_{micro} > 0$, we can express the value of information as:

$$EVPI_{micro} = \int_{-\infty}^{0} \left( -INMB_{micro} \frac{1}{\sqrt{Var(INMB_{micro})}} \varphi\left( \frac{INMB_{micro} - EINMB_{micro}}{\sqrt{Var(INMB_{micro})}} \right) \right) dINMB_{micro} \quad (23)$$

Likewise, in the case of $EINMB_{micro} < 0$, this expression is then:

$$EVPI_{micro} = \int_{0}^{\infty} \left( INMB_{micro} \frac{1}{\sqrt{Var(INMB_{micro})}} \varphi\left( \frac{INMB_{micro} - EINMB_{micro}}{\sqrt{Var(INMB_{micro})}} \right) \right) dINMB_{micro} \quad (24)$$

*Definition of Bias*

As noted in the Introduction, the expected outcomes from simulations with the iSTM provide estimates of corresponding outcomes from the mean-process represented by cSTM. We define bias:

$$Bias_{micro}(Outcome_{micro}) = E\big(E_\Theta(Outcome_{micro} - Outcome_{cSTM})\big) \quad (25)$$

In other words, the iSTM estimate of an expected outcome for the mean-process represented is biased if it differs systematically from that produced by the mean-process model (cSTM). It is possible, as we have noted above, that for some outcomes, the iSTM estimate may be unbiased ($Bias_{micro} = 0$) while for others, it may be biased. Furthermore, if for some outcome where the iSTM estimate is biased and the $Bias_{micro}$ approaches 0 as $n_{micro}$ approaches $\infty$, then the iSTM will be biased but consistent.

*Numerical Simulations*

To illustrate the results implied by the equations above, we perform numerical simulations. The goal is to show that there exist cases where the decision uncertainty and VOI outcomes generated by the iSTM are biased estimates of the outcomes from the corresponding cSTM and to show that the direction of bias for the decision uncertainty outcome can be either upwards or downwards.

First, we replicate outcomes from a cSTM accounting for parameter uncertainty by sampling from various distributions directly defined for the $INMB_{psa}$ (normal distribution or mixture of normal distributions [that might occur due to bimodal priors or heterogeneity]) for decision problems involving a given number of strategies (2 strategies or 3 strategies) which then implies either 1 or 2 INMB distributions relative to the comparator strategy. For each of these, to replicate outcomes from a corresponding iSTM, we then simulate samples from the corresponding distributions of $INMB_{micro}$ by adding normally distributed zero-mean noise to the samples of $INMB_{psa}$ where the variance of this zero-mean noise decreases with more microsimulated individuals (see the online appendix for further details). With these samples, we compute the EINMB, the probability of being cost-effective, and the EVPI for the cSTM and iSTM per the expressions above for these outcomes.

Next, we illustrate these results at various numbers of PSA samples and microsimulated individuals, and with various correlation structures for the incremental costs and incremental QALYs at the population and/or individual levels. We perform additional numerical simulations using the parameter values defined in Table 1. As Table 1 shows, there are, in fact, 9 sets of parameter values as we consider combinations of second-order correlation of incremental QALYs and incremental costs with treatment as well as how incremental QALYs and incremental costs may be correlated within microsimulated individuals (i.e., people who have higher than average incremental QALYs from treatment are also more likely to have higher (or lower) than average incremental costs with treatment). For each set of parameter values, we compute cSTM results with two sizes of PSA samples ($n_{psa} \in \{1000, 4000\}$). For each PSA sample size, we then compute iSTM results with different numbers of microsimulated individuals ($n_{micro} \in \{10, 40, 160, 640, 2560\}$). For each PSA sample size and number of microsimulated individuals, we bootstrap uncertainty intervals for the iSTM results by repeating the microsimulations 400 times and then taking the 2.5$^{th}$ and 97.5$^{th}$ percentile of the results.

To enable reproducibility, all code for the numerical simulations is included in the online appendix and persistent URL.

**Results**

Overall, in probabilistic analyses, while individual-level microsimulations produced unbiased estimates of EINMB when compared to equivalent cohort state transition Markov models, their estimates of decision uncertainty (e.g., the probability of being cost-effective at a given WTP threshold and EVPI) are systematically biased (but asymptotically consistent).

**Figure 1A** provides the underlying intuition. In the case of normally distributed INMB in a two-strategy decision problem: 1) the EINMB is the same for both the cSTM and iSTM; 2) there is greater variance in the INMB (and hence greater uncertainty in the estimate of the EINMB) with the iSTM due to the added Monte Carlo noise of microsimulating a finite number of individuals; 3) the iSTM estimate of the probability of not being cost-effective differs from the cSTM's estimate; and 4) the iSTMs overestimates EVPI compared to the cSTM. The misestimates and overestimates occur because while the distributions for the cSTM and iSTM are symmetric and have equal centers, statistics concerning decision uncertainty depend on one tail of the distribution and hence the differentially large variance of the iSTM's INMB distribution generates bias in its estimates of these statistics.

**Figure 1B** extends the intuition for a case where the INMB distribution is a mixture of normal distributions (i.e., not normally distributed). Like with the normally distributed INMB, the EINMB for the cSTM and iSTM are equal and the EVPI from the iSTM is overestimated. However, unlike the normally distributed case, the probability of being cost-effective is now underestimated with the iSTM. Hence, while this statistic is biased in both cases, the direction of bias is not obvious *a priori*.

*Biases in the case of 2 decision alternatives*

Our numerical simulations for the iSTM estimates of the EINMB show them to be unbiased estimates of those produced by a corresponding cSTM (**Figure 2A-2B**). This is the case regardless of whether incremental costs and incremental QALYs are correlated at the individual level or at the population level (**Appendix Figure 1**). One can increase the precision of an iSTM's estimate of the EINMB (i.e., reduce its variance) both by increasing the number of PSA samples and the number of individuals microsimulated. Notably, the estimates of the EINMB produced by an iSTM are less precise (i.e., have higher variance) with greater individual-level anti-correlation between incremental QALY gains with treatment and incremental costs with treatment (**Figure 2C**).

While iSTM estimates of EINMB are unbiased, their estimates of the probability of being cost-effective are biased (**Figure 3A**). In the case of a two alternative decision-problem and normally distributed incremental costs and QALYs, iSTM estimates are downwards biased when the cSTM estimate of the probability of being cost-effective is greater than 50% and upwards biased when the cSTM estimate is less than 50%. This is due to additional first-order Monte Carlo noise leading to increased variance, increasing the proportion of PSA samples for which the INMB is on the opposite below (above) 0 when treatment is the optimal (non-optimal) action. More generally, for decisions with larger numbers of alternatives and different distributions, patterns of bias can differ and are more complex (see below).

While the iSTM estimates of the probability of being cost-effective are biased, they are asymptotically consistent, in that as the number of microsimulated individuals is increased, the size of the bias is expected to decrease to 0. Importantly, the bias is not decreased by having more PSA samples because those represent second-order uncertainty. In addition, bias is larger when the within-person incremental QALYs with treatment and within-person incremental Costs with treatment are more negatively correlated (**Figure 3B**). **Appendix Figure 2** shows that there are some differences in the bias in the probability of being cost-effective due to correlation at the individual and population level in incremental QALYs and incremental Costs, but across all cases, the number of microsimulated individuals exerts an important influence on the size of this bias.

The iSTM estimates of EVPI are also biased (**Figure 4**). In our example, the direction of the bias is upwards. The estimates, while biased, are asymptotically consistent, in that as more individuals are microsimulated, the bias tends to 0. Since information has value when it allows one to shift from a sub-optimal decision to an optimal decision and since with iSTMs there is a larger fraction of PSA samples for which INMB estimates fall on the opposite side of 0 from the EINMB, a greater proportion of PSA samples will appear to have a positive value of information. Furthermore, since the variance of the INMB estimates with the iSTM is greater, there will be a higher proportion of samples with larger magnitude values of information. As can be seen in **Appendix Figure 3**, especially at lower WTP thresholds (e.g., below ~$12,500/QALY gained in the example), the bias in the EVPI is larger when the within-person incremental QALYs with treatment and within-person incremental costs with treatment are more negatively correlated. **Appendix Figure 3** shows that there are some differences in the EVPI bias due to correlation at the individual and population level in incremental QALYs and incremental Costs, but across all cases, the number of microsimulated individuals exerts an important influence on the size of this bias.

*Bias in the case of 3+ decision alternatives*

We examine the case of a decision problem with 3 alternatives where the INMBs for two of the alternatives (Strategies 1 and 2) are defined with respect to the status quo (Strategy 0). To illustrate some of the potentially complex patterns of bias in the probability of cost-effectiveness and EVPI, we simulate many cases of this decision problem which differ in terms of the means and variances of the INMB distributions (normal) defined for the alternatives and the amount of 0-mean Monte Carlo noise that exists for each alternative when using an iSTM to simulate it. In all cases, Strategy 1 has the highest EINMB and is therefore the preferred option with current information. Strategy 2 has a lower EINMB, which can either be positive (better than the status quo Strategy 0) or negative (worse than the status quo). In total, we simulate 5,760 different combinations of EINMBs and amount of Monte Carlo noise in the INMB distributions due to microsimulation (**Appendix Table 1**).

**Figure 5** shows patterns of bias for three illustrative cases. Given the normality assumption for the INMB distributions, the probability that Strategy 1 (the preferred strategy on expectation) is optimal computed with the iSTM is downwards biased in these cases. Because of this, the probability of another alternative being optimal is overestimated for at least one of the two other alternatives. However, the cases show that sometimes the probabilities for Strategies 0 and 2 can be both overestimated, but at other times, for one alternative, the probability is overestimated and for the other, the probability is underestimated. For these cases, just as with the 2-alternative decision problems considered above, the bias in EVPI estimated with the iSTM is ≥0 (upwards biased) (see **Appendix Figures 4-6** for full details of the patterns of bias for the probabilities of being cost-effective and the EVPI for the 3,840 combinations we simulated).

**Discussion**

Economic evaluation in health and medicine is witnessing three converging trends. The first is a shift from cohort state transition models (i.e., Markov cohort models, cSTMs) to individual-level state transition models (i.e., microsimulation or iSTMs) to enable modelers to reflect more complex and heterogeneous diseases and intervention strategies. The second is the use of probabilistic analyses to appropriately estimate the expectations of outcomes simulated using these decision models and to propagate population-level parameter uncertainty in these expectations. The third is a growing interest in gauging current levels of decision uncertainty and the value of collecting additional information to resolve that uncertainty.

Given these trends, our study considers the potential for bias introduced into estimates of decision uncertainty and value of information through iSTMs whose outputs also necessarily contain first-order Monte Carlo noise. We confirm both analytically and through a range of numerical simulations for iSTMs that while their estimates of the expected incremental value of decision alternatives are unbiased with respect to corresponding cSTMs, their estimates of decision uncertainty and the value of information are biased but asymptotically consistent.

A key implication of the study findings involves how analysts conducting economic evaluations with iSTMs allocate their computational budgets.[21-23] For computing expected incremental net monetary benefit, the number of parameter sets sampled from the parameter uncertainty distributions (i.e., outer loops) and the number of microsimulated individuals simulated for each parameter set sampled (i.e., inner loops) are interchangeable. However, for decision uncertainty and value of information, taking more samples from the parameter uncertainty distribution does not reduce the bias in the estimates produced by the iSTM; only simulating larger numbers of microsimulated individuals will reduce this bias. Hence, the appropriate balance between inner and outer loop samples depends on the quantities that the analyst is interested in estimating and the precision required for each of them requiring trade-offs in their accuracy given the computation budget.

Given the study findings, one area of future research is developing efficient diagnostics to gauge the magnitude of the bias for given combinations of inner and outer loop samples, which can be used to suggest efficient computational budget allocations. For example, while more samples are better if one ignores computational budgets, how many samples would be sufficient to make the iSTM bias suitably small depends on the model used and the decision being considered. A model focused on comparing two highly effective strategies for preventing a relatively rare disease outcome (e.g., two aggressive screening strategies for a relatively rare cancer) may have to simulate many more individuals to form stable estimates of incremental net benefits of one strategy versus the other than if the outcome were common or the strategies being compared were substantially different in their costs and the health benefits they produced. Approaches might use a fraction of the overall computational budget to perform a probabilistic analysis with a given number of outer (e.g., 1,000) and inner loops (e.g., 400) and then act as if all of these samples were not available to consider how the model estimates as if there were 1,000 outer and 200 inner samples or 1,000 outer and 1,000 inner samples. One could then use a form of convergence analysis to examine how and at what rate the estimates changed as a function of the number of inner loop samples. In principle, one could adapt value of information approaches concerning optimal sample size and study design to this situation as well.[27]

While our study has established that there exist cases where iSTMs generate biased but asymptotically consistent estimates of decision uncertainty and value of information quantities with respect to equivalent cSTMs, the characterization of the situations in which these biases are most pronounced and most consequential will require additional work. One important extension is greater consideration of various distributions of incremental net monetary benefit as well as the distribution of the incremental cost and incremental health benefits and their correlations within and between the population and individual levels. Another important extension is considering these biases as the number of interventions increases. A third extension would be to consider the biases with iSTMs for estimating other measures of value of information like partial perfect information and the expected net benefit of sampling and optimal sample size. A fourth area would be to gauge the potential size of biases in the existing applied literature by examining the estimates of these quantities from published model-based analyses using iSTMs and reimplementing the equivalent cSTMs to calculate the biases and perhaps re-performing the iSTM analyses with different inner and outer loop sample sizes. In summary, there are numerous opportunities for promising future work in this area.

Our study is careful to define bias as a systematic deviation in estimating decision-analytic outcomes of the mean process (cSTM) using a corresponding iSTM. The mean-process outcomes are those relevant to a risk-neutral decision maker and in situations where the actual population for whom the decision is relevant is large. In this case, the additional noise from microsimulation is essentially a nuisance that should be reduced and removed to the extent possible.[28-30] However, in the case of risk-averse decision makers in situations where the population for whom decision is relevant is small (i.e., so that the number of microsimulated individuals might be equal to the number of actual people in that population itself), the noise from the iSTM may be relevant. For example, if one were deciding between two alternative actions for an institution housing 100 people, the fraction of time that one alternative action will produce the highest INMB should include Monte Carlo noise, although such an estimate systematically differs from what one would have estimated for a very large institution using a cSTM. In other words, in these situations, the iSTM is not introducing bias but rather telling us something very important that we may well want to know in addition to what mean-process estimates might tell us.

In conclusion, despite being unbiased in identifying the best strategy on expectation, iSTMs introduce bias in estimates of decision uncertainty and value of information. Fortunately, as the bias is asymptotically consistent, the size of the bias can be reduced by using sufficiently large numbers of microsimulated individuals per probabilistic sample. Analysts using iSTMs for economic evaluations that include probabilistic analyses should explicitly demonstrate that the number of microsimulated individuals per probabilistic sample is sufficiently large to make the bias in these estimates small.

**Acknowledgments**

**Statements and Declarations**

**References**
1. Neumann, Peter J., and others (eds), Cost-Effectiveness in Health and Medicine, 2nd edn (New York, 2016; online edn, Oxford Academic, 17 Nov. 2016), https://doi.org/10.1093/acprof:oso/9780190492939.001.0001, accessed 2 Sept. 2024.
2. NICE health technology evaluations: the manual (PMG36). https://www.nice.org.uk/process/pmg36/resources/nice-health-technology-evaluations-the-manual-pdf-72286779244741, accessed 2 Sept. 2024.
3. Guidelines for the Economic Evaluation of Health Technologies: Canada — 4th Edition. https://www.cda-amc.ca/guidelines-economic-evaluation-health-technologies-canada-4th-edition, accessed 2 Sept. 2024.
4. Briggs AH, Weinstein MC, Fenwick EA, Karnon J, Sculpher MJ, Paltiel AD; ISPOR-SMDM Modeling Good Research Practices Task Force. Model parameter estimation and uncertainty analysis: a report of the ISPOR-SMDM Modeling Good Research Practices Task Force Working Group-6. Med Decis Making. 2012 Sep-Oct;32(5):722-32.
5. Goldhaber-Fiebert JD, Stout NK, Ortendahl J, Kuntz KM, Goldie SJ, Salomon JA. Modeling human papillomavirus and cervical cancer in the United States for analyses of screening and vaccination. Popul Health Metr. 2007 Oct 29;5:11.
6. Goldhaber-Fiebert JD, Brandeau ML. Modeling and calibration for exposure to time-varying, modifiable risk factors: the example of smoking behavior in India. Med Decis Making. 2015 Feb;35(2):196-210.
7. Goldhaber-Fiebert JD, Jalal HJ. Some Health States Are Better Than Others: Using Health State Rank Order to Improve Probabilistic Analyses. Med Decis Making. 2016 Nov;36(8):927-40.
8. Alarid-Escudero F, MacLehose RF, Peralta Y, Kuntz KM, Enns EA. Nonidentifiability in Model Calibration and Implications for Medical Decision Making. Med Decis Making. 2018 Oct;38(7):810-821.
9. Jalal H, Trikalinos TA, Alarid-Escudero F. BayCANN: Streamlining Bayesian Calibration With Artificial Neural Network Metamodeling. Front Physiol. 2021 May 25;12:662314.
10. Alarid-Escudero F, Knudsen AB, Ozik J, Collier N, Kuntz KM. Characterization and Valuation of the Uncertainty of Calibrated Parameters in Microsimulation Decision Models. Front Physiol. 2022 May 9;13:780917.
11. Chrysanthopoulou SA, Rutter CM, Gatsonis CA. Bayesian versus Empirical Calibration of Microsimulation Models: A Comparative Analysis. Med Decis Making. 2021 Aug;41(6):714-726.
12. Krijkamp EM, Alarid-Escudero F, Enns EA, Jalal HJ, Hunink MGM, Pechlivanoglou P. Microsimulation Modeling for Health Decision Sciences Using R: A Tutorial. Med Decis Making. 2018 Apr;38(3):400-422.
13. Mertens E, Genbrugge E, Ocira J, Peñalvo JL. Microsimulation Modeling in Food Policy: A Scoping Review of Methodological Aspects. Adv Nutr. 2022 Mar;13(2):621-632.
14. Iskandar R. Adding noise to Markov cohort state-transition model in decision modeling and cost-effectiveness analysis. Stat Med. 2020 May 15;39(10):1529-1540.
15. Basu A, Meltzer D. Value of information on preference heterogeneity and individualized care. Med Decis Making. 2007 Mar-Apr;27(2):112-27.
16. Kasztura M, Richard A, Bempong NE, Loncar D, Flahault A. Cost-effectiveness of precision medicine: a scoping review. Int J Public Health. 2019 Dec;64(9):1261-1271.
17. Pitt AL, Goldhaber-Fiebert JD, Brandeau ML. Public Health Interventions with Harms and Benefits: A Graphical Framework for Evaluating Tradeoffs. Medical Decision Making. 2020;40(8):978-989.
18. Podolsky MI, Present I, Neumann PJ, Kim DD. A Systematic Review of Economic Evaluations of COVID-19 Interventions: Considerations of Non-Health Impacts and Distributional Issues. Value Health. 2022 Aug;25(8):1298-1306.

19. Avanceña ALV, Prosser LA. Examining Equity Effects of Health Interventions in Cost-Effectiveness Analysis: A Systematic Review. Value Health. 2021 Jan;24(1):136-143.
20. Kuntz, Karen M, and Milton C Weinstein, 'Modelling in economic evaluation', Economic Evaluation in Health Care: Merging theory with practice (New York, NY, 2001; online edn, Oxford Academic, 31 Oct. 2023), https://doi.org/10.1093/oso/9780192631770.003.0007, accessed 2 Sept. 2024
21. O'Hagan A, Stevenson M, Madan J. Monte Carlo probabilistic sensitivity analysis for patient level simulation models: efficient estimation of mean and variance using ANOVA. Health Econ. 2007 Oct;16(10):1009-23.
22. Hatswell AJ, Bullement A, Briggs A, Paulden M, Stevenson MD. Probabilistic Sensitivity Analysis in Cost-Effectiveness Models: Determining Model Convergence in Cohort Models. Pharmacoeconomics. 2018 Dec;36(12):1421-1426.
23. Yaesoubi R. How Many Monte Carlo Samples are Needed for Probabilistic Cost-Effectiveness Analyses? Value Health. 2024 Jul 6:S1098-3015(24)02755-4.
24. Lee TY, Gustafson P, Sadatsafavi M. Closed-Form Solution of the Unit Normal Loss Integral in 2 Dimensions, with Applications in Value-of-Information Analysis. MDM. 2023 Jul: 43(5): 621-626.
25. Schlaifer RO. Probability and Statistics for Business Decisions: An Introduction to Managerial Economics Under Uncertainty. New York: McGraw-Hill; 1959.
26. Jalal H, Goldhaber-Fiebert JD, Kuntz KM. Computing expected value of information from probabilistic sensitivity analysis using linear regression metamodeling. MDM. 2015; 35(5):584-95.
27. Conti S, Claxton K. Dimensions of design space: a decision-theoretic approach to optimal research design. Med Decis Making. 2009 Nov-Dec;29(6):643-60.
28. Reitsma MB, Salomon JA, Goldhaber-Fiebert JD. Common Random Numbers for Stochastic Network-Based Transmission Dynamic Models. (Abstract: #23940 [QMTD]). Presented to the 2023 Society of Medical Decision Making Annual Meeting, Philadelphia, PA.
29. Li Z, Knowlton G, Wheatley MM, Jenness SM, Enns EA. Linear Regression Metamodeling for Variance Reduction in Probabilistic Sensitivity Analysis. (Abstract: #23897 [QMTD]). Presented to the 2023 Society of Medical Decision Making Annual Meeting, Philadelphia, PA.
30. Stout NK, Goldie SJ. Keeping the noise down: common random numbers for disease simulation modeling. Health Care Manag Sci. 2008 Dec;11(4):399-406.

**Table 1. Parameter values for numerical simulations**

| Parameter | Description | Value for Example |
|---|---|---|
| $\Delta Q_{psa}$ | Mean incremental QALYs with treatment for second-order uncertainty distribution | 0.02 |
| $\Delta C_{psa}$ | Mean incremental Costs with treatment for second-order uncertainty distribution | 250 |
| $\sigma_{\Delta Q}$ | Standard Deviation of Incremental QALYs with treatment for second-order uncertainty distribution | 0.01 |
| $\sigma_{\Delta C}$ | Standard Deviation of Incremental Costs with treatment for second-order uncertainty distribution | 20 |
| $\rho_{psa}$ | Correlation in the second-order uncertainty of incremental QALYs and incremental Costs* | {-0.9, 0, 0.9} |
| $\sigma_{\Delta Q, micro}$ | Standard Deviation of incremental QALYs with treatment across individuals in the microsimulation | 0.20 |
| $\sigma_{\Delta C, micro}$ | Standard Deviation of incremental Costs with treatment across individuals in the microsimulation | 400 |
| $\rho_{micro}$ | Correlation within individuals of incremental QALYs and incremental Costs* | {-0.9, 0, 0.9} |

* Correlations show multiple values as we conduct numerical simulations with all combinations of positive, negative, and no correlations in the second-order uncertainty distribution and the within-person outcomes.

**Figure 1. Comparison of outcomes for a two-alternative decision problem evaluated with an iSTM and a corresponding cSTM**

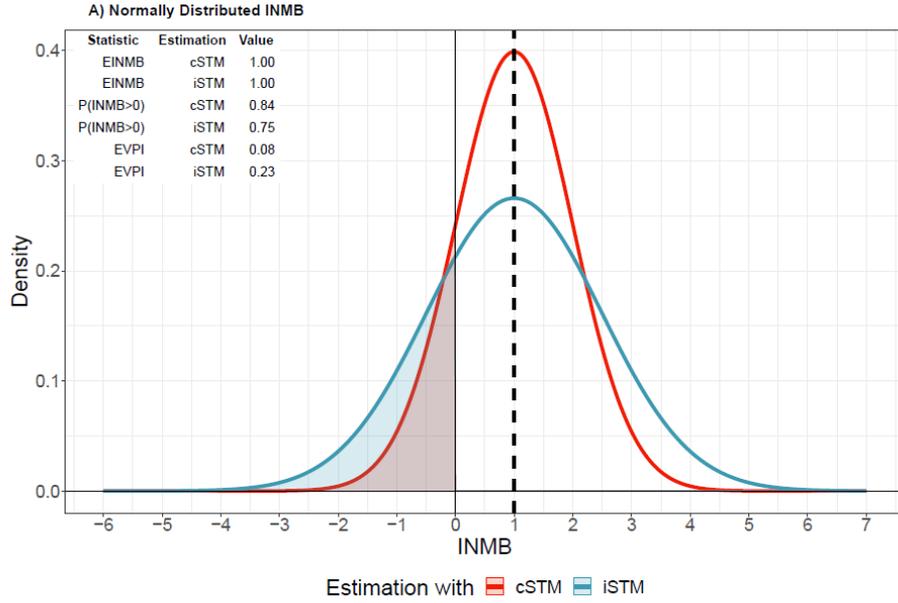

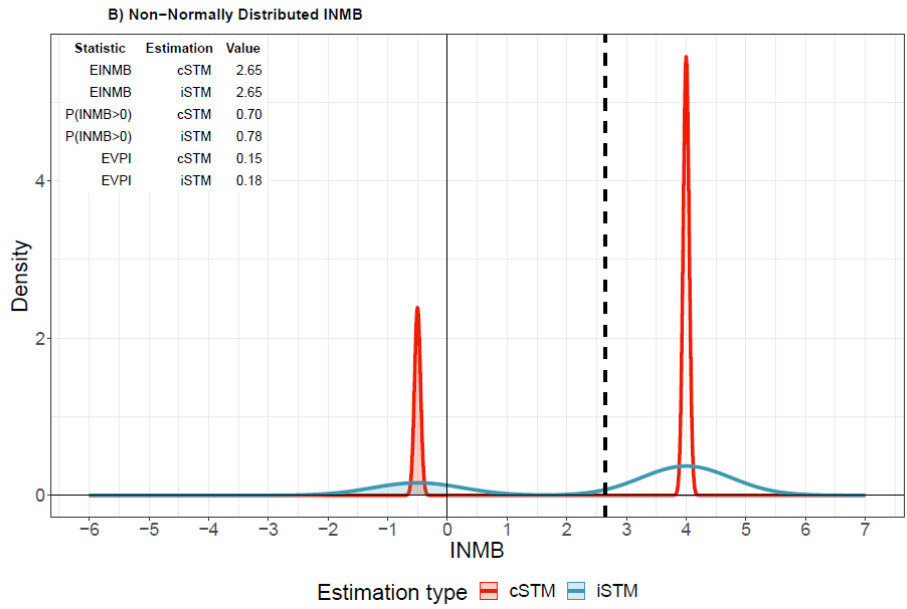

**Figure 2. iSTMs are Unbiased Estimators of the EINMB with Lower Precision when the Person-Level Uncertainties in Incremental QALYs and Incremental Costs have more Negative Correlation**

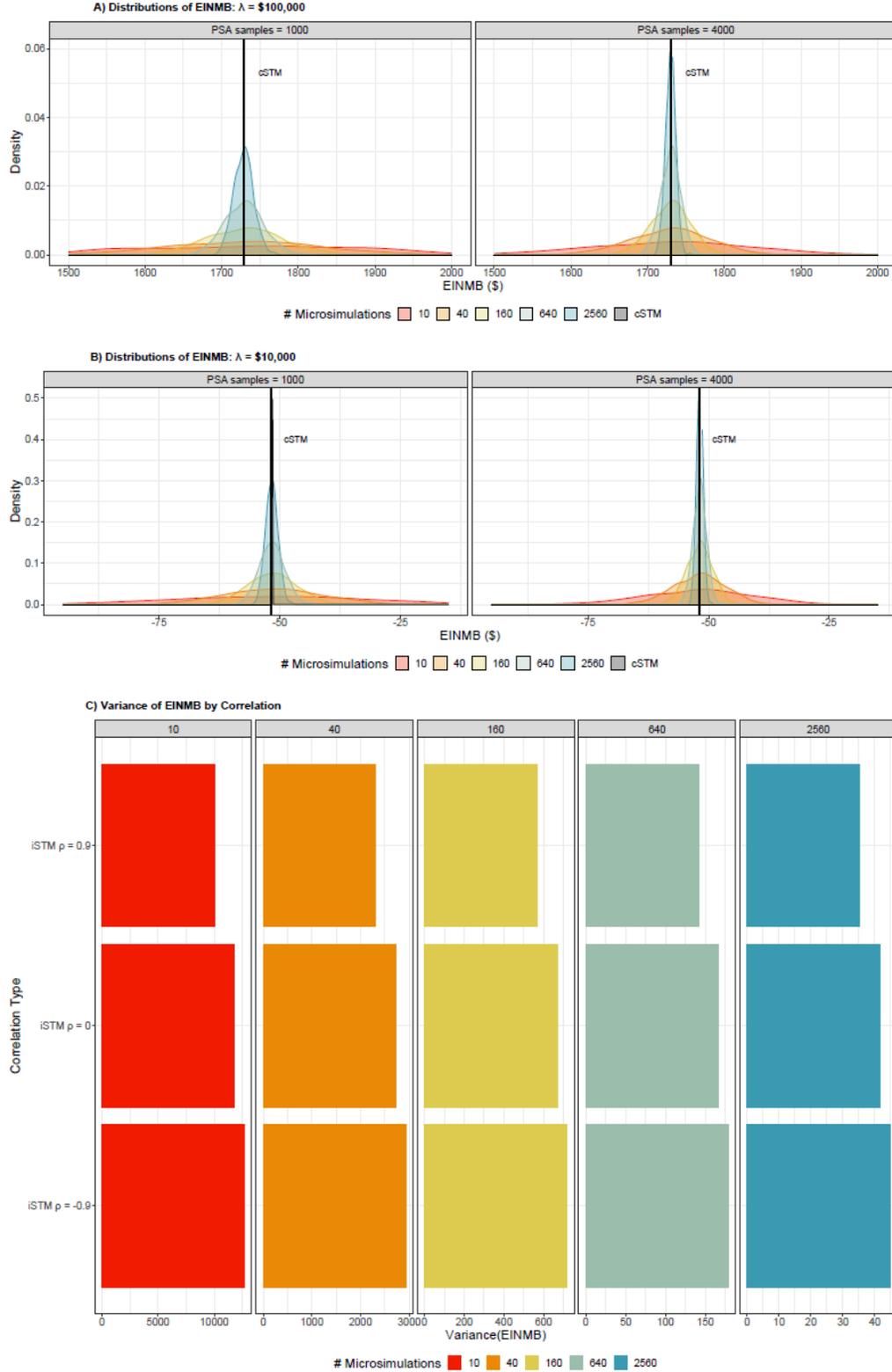

**Figure 3. iSTMs are Biased but Consistent Estimators of the Probability that a Strategy is Cost-Effective with Larger Bias when the Person-Level Uncertainties in Incremental QALYs and Incremental Costs have More Negative Correlation**

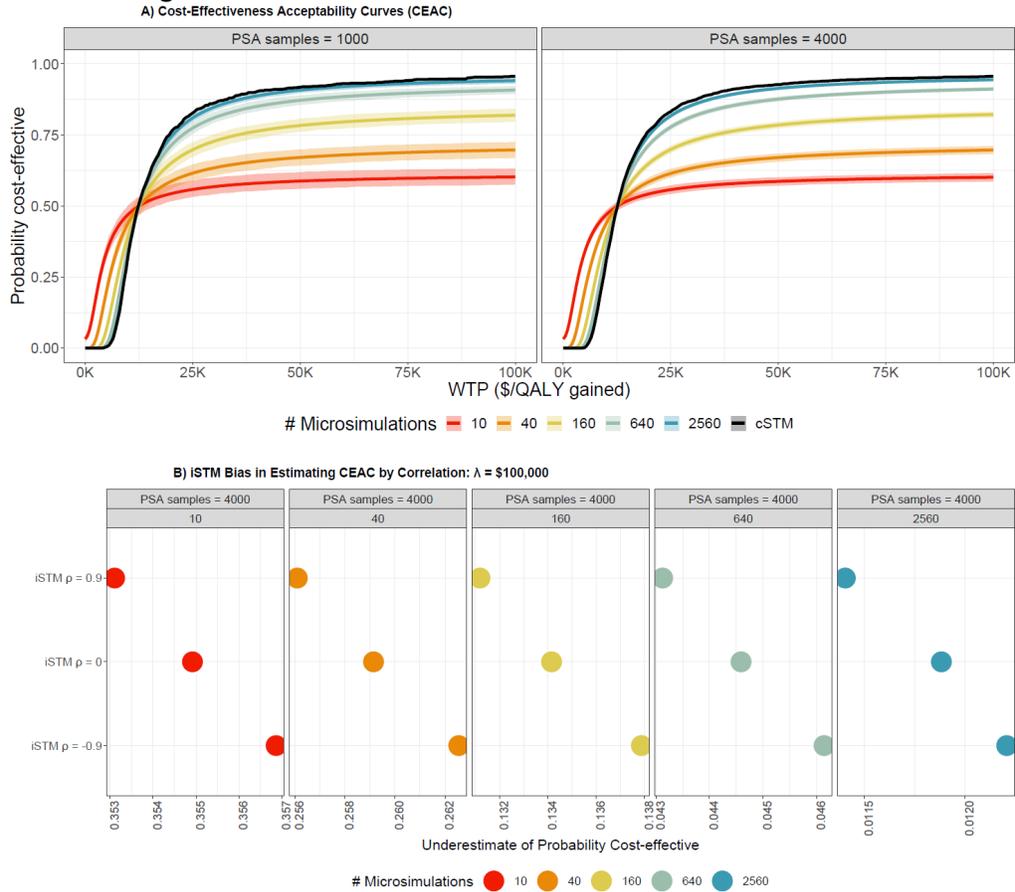

**Figure 4. iSTMs are Biased but Consistent Estimators of EVPI**

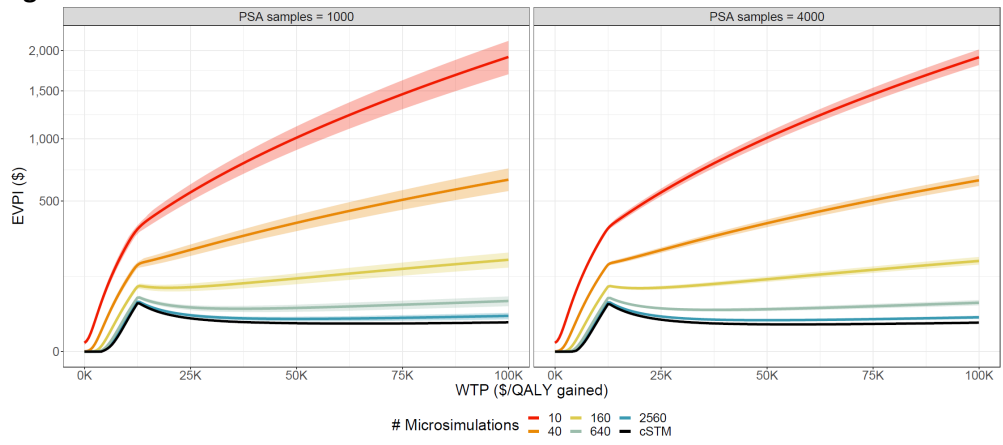

**Figure 5. iSTMs' Direction of Bias for Estimating the Probability of Being Cost-Effective is Complex When Considering 3 or More Strategies Even with Normally Distributed INMBs**

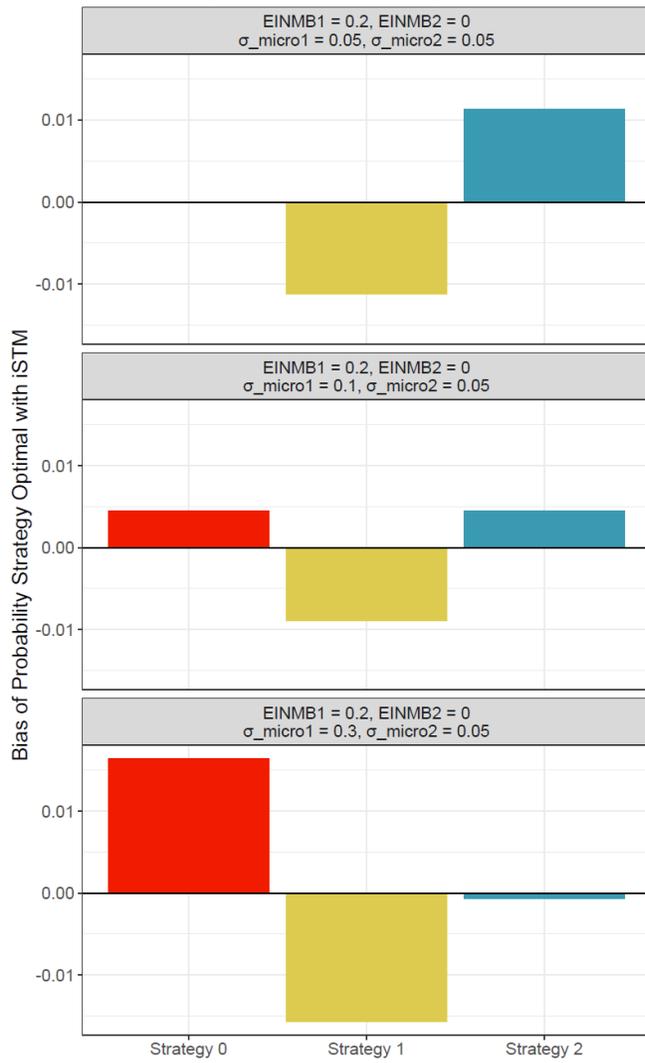

**Supplemental Appendix:**
**Microsimulation Estimates of Decision Uncertainty and Value of Information Are Biased but Consistent**

Jeremy D. Goldhaber-Fiebert, PhD (1,2), Hawre Jalal, MD, PhD (3), Fernando Alarid Escudero, PhD (1,2)

(1) Department of Health Policy, Stanford School of Medicine, Stanford, CA, USA
(2) Center for Health Policy, Freeman Spogli Institute, Stanford University, Stanford, CA, USA
(3) School of Epidemiology and Public Health, University of Ottawa, Ottawa, Ontario, Canada

**Appendix Figure 1. EINMB estimated with iSTMs across WTP Thresholds, Population-Level Correlation in the Uncertainties of Incremental QALYs and Incremental Costs, and Individual-Level Correlation in the Uncertainties of Incremental QALYs and Incremental Costs**

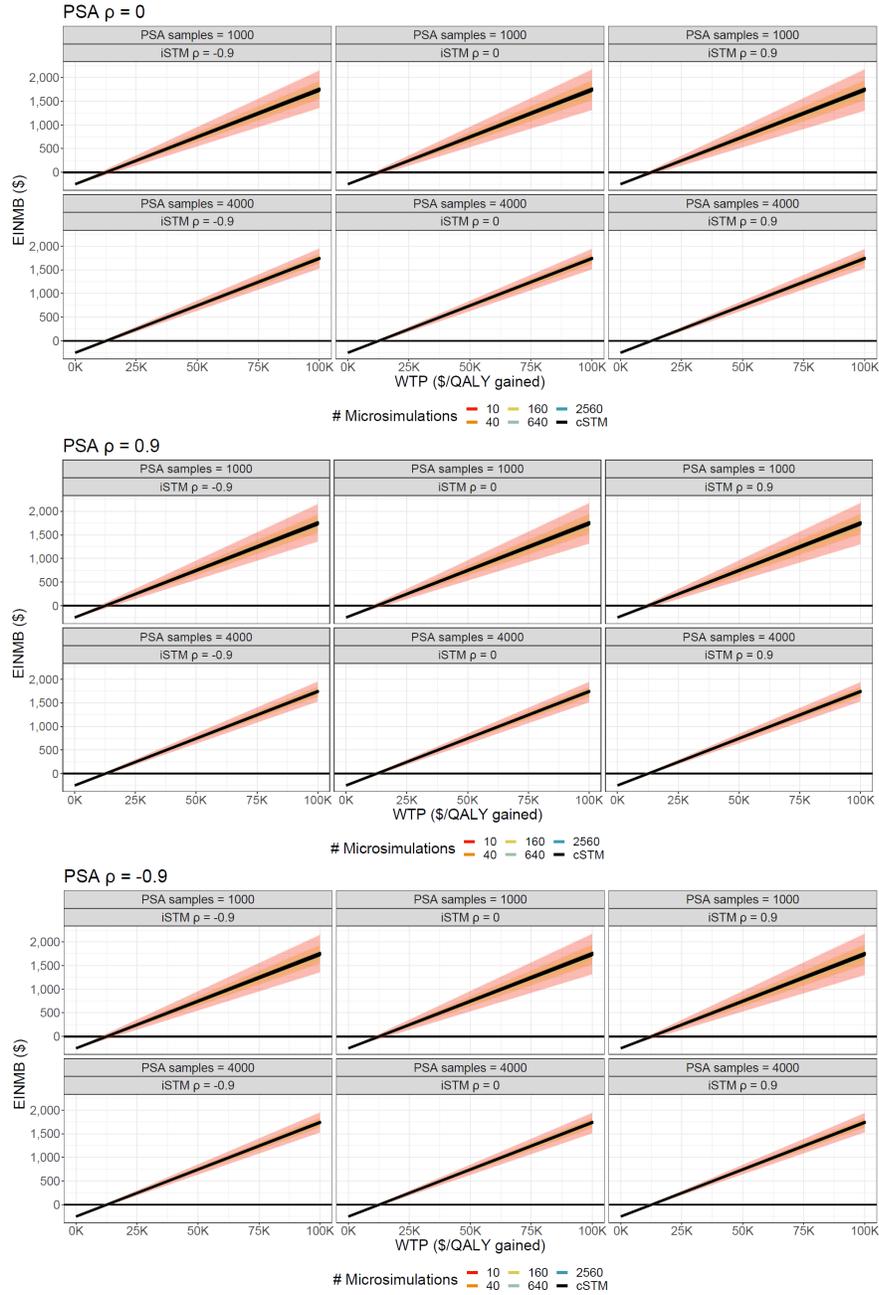

**Appendix Figure 2. Probability of Being Cost-Effective estimated with iSTMs across WTP Thresholds, Population-Level Correlation in the Uncertainties of Incremental QALYs and Incremental Costs, and Individual-Level Correlation in the Uncertainties of Incremental QALYs and Incremental Costs**

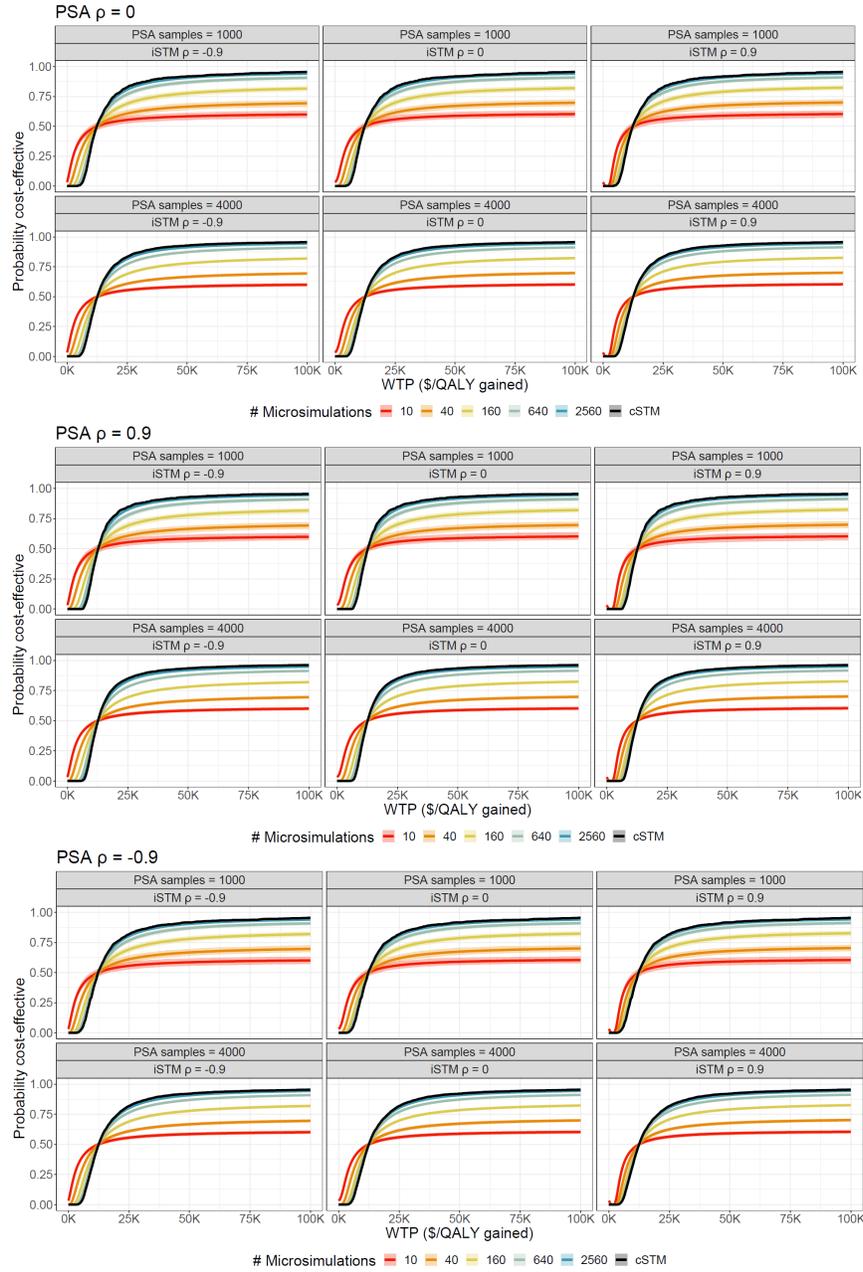

**Appendix Figure 3. EVPI estimated with iSTMs across WTP Thresholds, Population-Level Correlation in the Uncertainties of Incremental QALYs and Incremental Costs, and Individual-Level Correlation in the Uncertainties of Incremental QALYs and Incremental Costs**

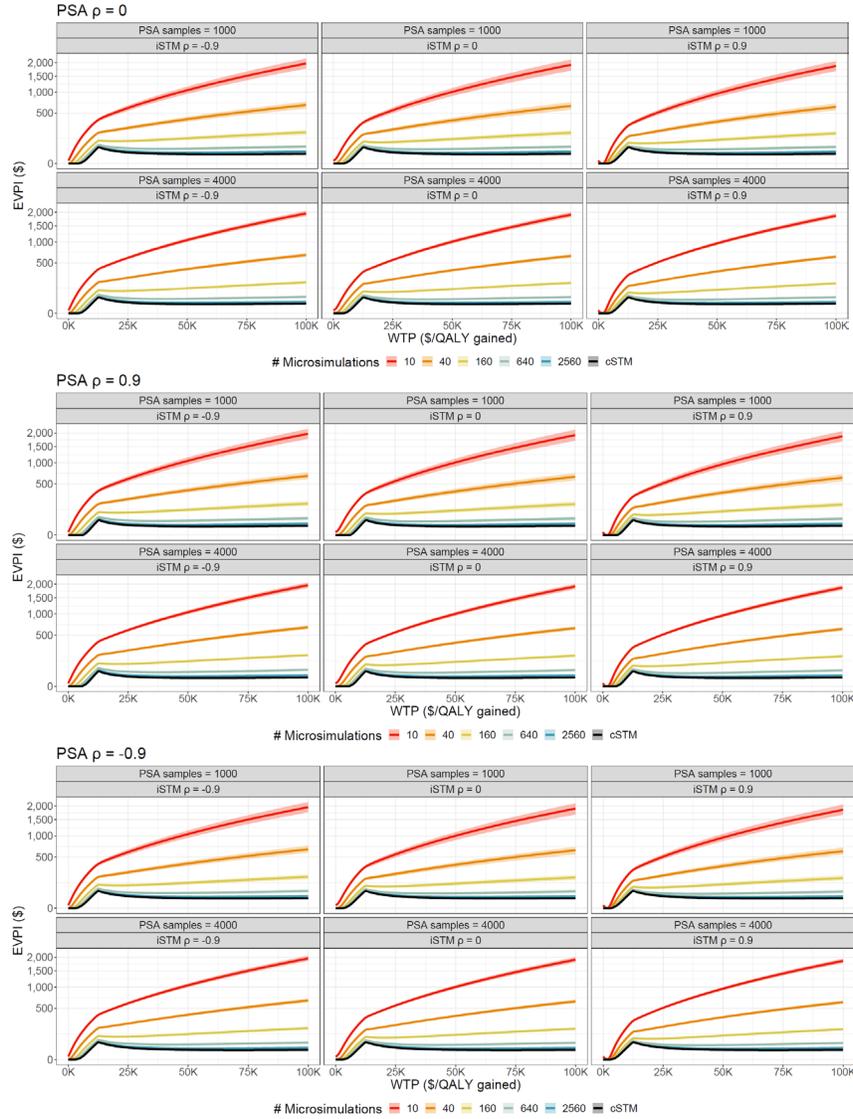

We perform simulations of our 3-strategy decision problem example using combinations of parameters from Appendix Table 1 below (5,760 combinations in total).

**Appendix Table 1. Parameter values for numerical simulations for 3-strategy example**

| Parameter | Description | Value for Example |
|---|---|---|
| $\text{EINMB}_1$ | EINMB of Strategy 1 | {0.20, 0.40, 0.60} |
| $decrement\text{EINMB}_2$ | How much lower the EINMB of Strategy 2 is compared to Strategy 1's EINMB | {0.10, 0.15, 0.20, 0.25, 0.30, 0.50} |
| $\sigma_{\text{INMB1,psa}}$ | Standard Deviation of the INMB of Strategy 1 | {0.10, 0.20, 0.30, 0.40} |
| $\sigma_{\text{INMB2,psa}}$ | Standard Deviation of the INMB of Strategy 2 | {0.01, 0.05, 0.10, 0.20, 0.30} |
| $\sigma_{\text{INMB1,micro}}$ | Standard Deviation of the mean-0 noise added to the INMB distribution of Strategy 1 due to finite number of individuals in the microsimulations | {0.05, 0.10, 0.15, 0.30} |
| $\sigma_{\text{INMB2,micro}}$ | Standard Deviation of the mean-0 noise added to the INMB distribution of Strategy 2 due to finite number of individuals in the microsimulations | {0.05, 0.10, 0.15, 0.30} |

**Appendix Figure 4. The Magnitude and Direction of Bias in the Probability of Being Optimal for Each Strategy in the 3-Strategy Example Depends on the EINMBs (relative to Strategy 0) (μ1 and μ2) of Two Alternative Strategies (Strategies 1 and 2) and the Magnitudes of the Stochastic Noise in the INMBs (σ1 and σ2) of the Two Alternative Strategies**

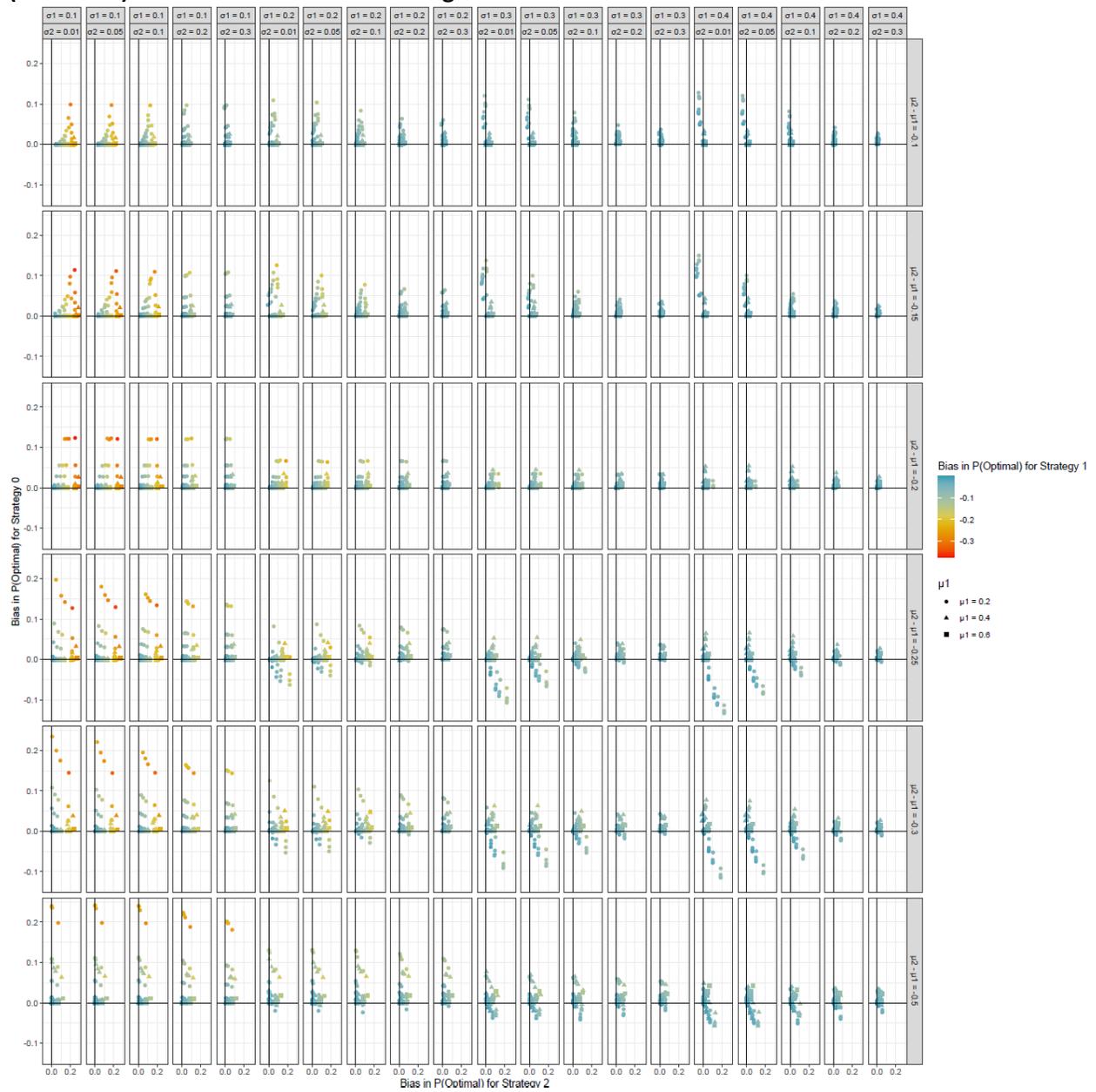

**Appendix Figure 5. The Magnitude and Direction of Bias in the Probability of Being Optimal for Each Strategy in the 3-Strategy Example Are Larger (all else equal) When the EINMB of the Strategy Preferred on Expectation is Larger/Farther Away from the EINMBs of the Other Two Strategies**

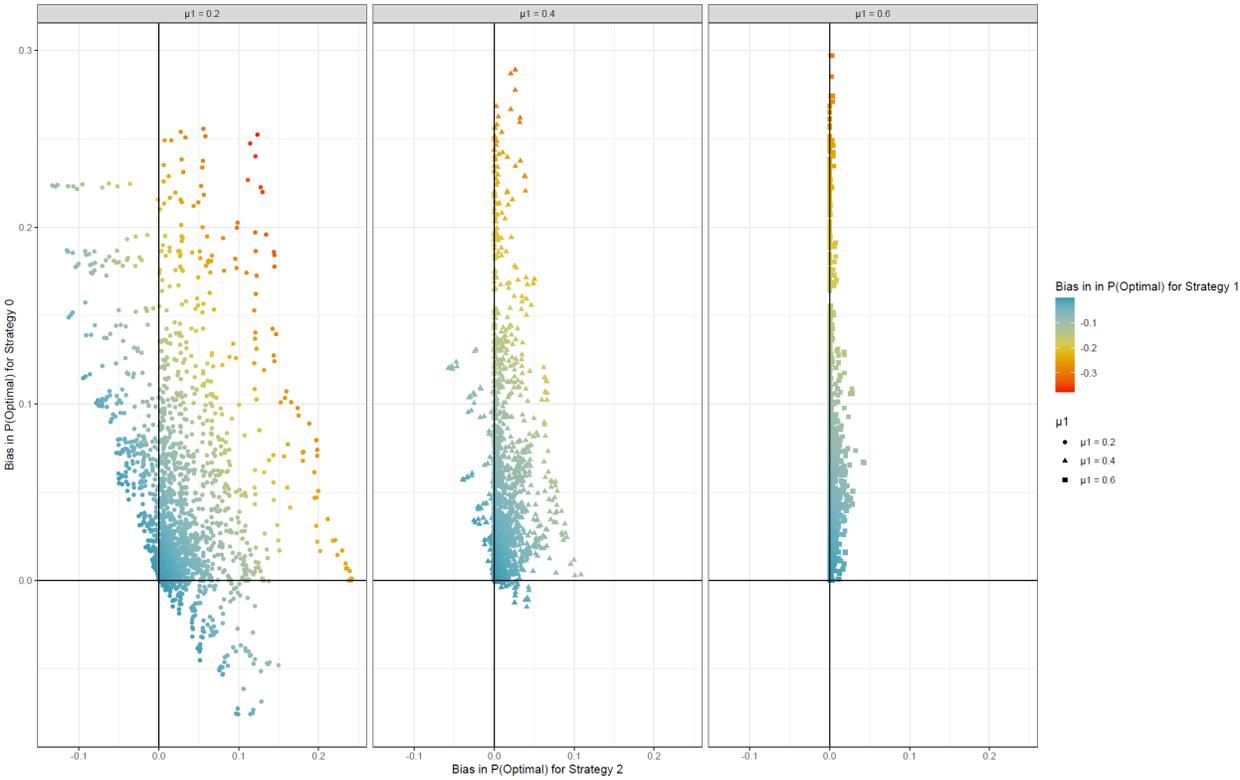

**Appendix Figure 6. The Magnitude of Bias in EVPI in the 3-Strategy Example is Larger (all else equal) When the Bias in the Probability of Being Cost-Effective for the Strategy Preferred on Expectation (Strategy 1) is Larger**

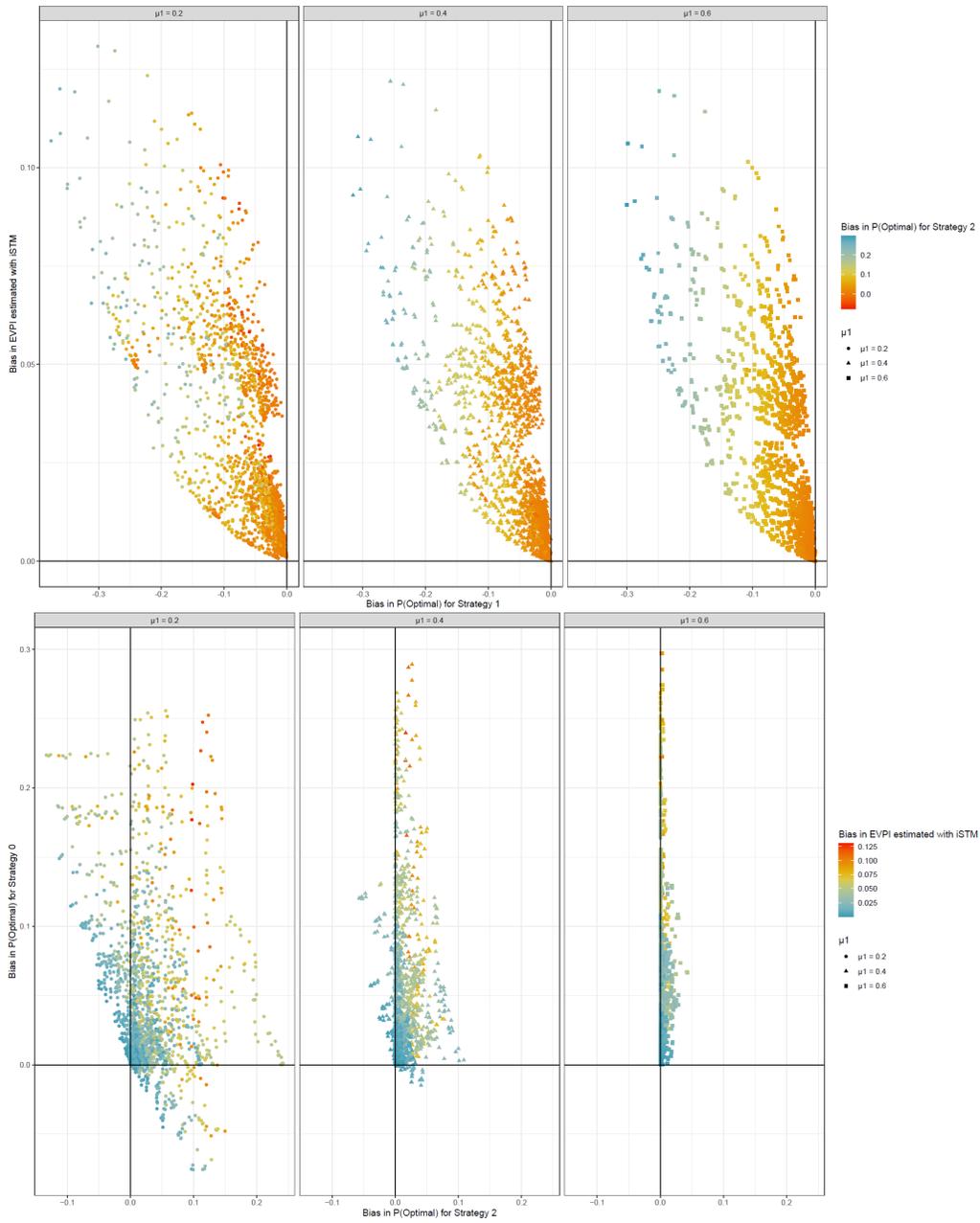